\documentclass[preprint,showpacs,preprintnumbers,amsmath,amssymb]{revtex4}
\usepackage{graphicx}
\usepackage{dcolumn}
\usepackage{bm}
\usepackage{color}

\newtheorem{remark}{Remark}

\def \rmd{\mathrm{d}}
\def \rme{\mathrm{e}}

\begin{document}
\title[Self-organization in foliated phase space]
{Self-organization in foliated phase space: 
\\
Construction of a scale hierarchy by adiabatic invariants of magnetized particles}
\author{Z. Yoshida$^1$ and S. M. Mahajan$^2$}
\affiliation{
$^1$Graduate School of Frontier Sciences, The University of Tokyo,
Kashiwa, Chiba 277-8561, Japan
\\
$^2$Institute for Fusion Studies, The University of Texas at Austin,
Austin, Texas 78712, U.S.A.
}

\begin{abstract}%
Adiabatic invariants foliate phase space, and
impart a macro-scale hierarchy by separating microscopic variables.
On a macroscopic leaf,
long-scale ordered structures are created while maximizing entropy.
A plasma confined in a magnetosphere is invoked to unveil the organizing principle
---in the vicinity of a magnetic dipole,
the plasma self-organizes to a state with a steep density gradient. 
The resulting nontrivial structure has maximum entropy in an appropriate, constrained phase space. 
One could view such a phase space as a leaf foliated in terms of Casimir invariants 
---adiabatic invariants measuring the number of quasi-particles (macroscopic representation of periodic motions) are identified as the relevant Casimir invariants.
The density clump is created in response to the inhomogeneity of the energy levels (frequencies) of the quasi-particles.
\end{abstract}


\maketitle

\section{Introduction}
The process of self-organization of a structure may appear to be an antithesis of the maximum entropy \emph{ansatz},
yet various nonlinear systems display what may be viewed as the simultaneous existence of order and disorder.
This co-existence begins to make sense if the self-organization processes and the entropy principle
were to manifest on different scales; 
disorder can still develop at a microscopic scale while an ordered structure emerges on some appropriate macroscopic scale. 
Writing a theory of self-organization, then, will be an exercise in delineating and understanding the characteristic \emph{scale hierarchy} of the physical system.

A biological body is a typical example in which an evident hierarchical structure is preprogrammed
enabling effective consumption of energy and materials as well as emission of entropy and waste. 
A physical macro-system ---a collective system of simple elements (a gravitational system, a plasma, etc.)--- 
is anchored on a different framework. 
An automatic emergence of scale hierarchy is not programmed; 
yet the controlling nonlinear dynamics can mimic a fundamental process of \emph{creation}. 

In this paper we develop a new framework to expose the creation process in action.  
The ordering principle is generally epitomized in a constraint ---a possible conservation law--- that, by restricting the class of motions available to the system, limits its ability to degenerate into general disorder.
The \emph{effective} phase space (ensemble) limited by such a constraint
is the relevant \emph{macro-hierarchy} on which nontrivial structures emerge.
By invoking the geometrical notion of Hamiltonian mechanics,
we formulate a macro-hierarchy as a \emph{Casimir leaf} of foliated phase space,
i.e., the level-set of a \emph{Casimir invariant}\,\cite{morrison98}.
The connection between the notion of scale hierarchy and a Casimir invariant
(an \emph{a priori} geometrical structure of the phase space) 
is built by identifying a Casimir invariant as an \emph{adiabatic invariant};
the adiabaticity criterion, then, determines what is macro.
After the microscopic action is separated, 
the macroscopic object, which we call a \emph{quasi-particle}, 
resides on a Casimir leaf.
We will construct the Boltzmann distribution of quasi-particles on a Casimir leaf.
The Casimir invariant represents the number of quasi-particles, which is the
determinant of the corresponding \emph{grand canonical ensemble}.
Interestingly, heterogeneity (structure) is created by the distortion
of the metric (invariant measure) dictating \emph{equipartition} on the leaf.

Basic physical mechanisms and processes,
that embodies our general framework of describing macro-hierarchy and self-organization,
will be  brought to light via investigating a magnetospheric plasma.
Magnetospheric plasmas
(the naturally occurring ones such as the planetary magnetospheres\,\cite{schulz1974,brautigam2000,Chen2007},
as well as their laboratory simulations\,\cite{boxer2010,yoshida2010,saito2010,saito2011})
are self-organized around the dipole magnetic fields 
in which charged particles cause a variety of interesting phenomena:
the often observed \emph{inward diffusion} (or up-hill diffusion)
of particles injected from the outer region is of particular interest.
This process is driven by some spontaneous fluctuations (symmetry breaking) 
that violate the constancy of angular momentum.
In a strong enough magnetic field, the canonical angular momentum $P_\theta$
is dominated by the magnetic part $q \psi$: the charge multiplied by the flux function
(in the $r$-$\theta$-$z$ cylindrical coordinates,
$\psi=rA_\theta$, where $A_\theta$ is the $\theta$ component of the vector potential).
The conservation of $P_\theta\approx q\psi$, therefore, restricts the particle motion
to the magnetic surface (level-set of $\psi$). 
It is only via randomly-phased fluctuations that the particles can diffuse across magnetic surfaces. 
Although the diffusion is normally a process  that diminishes gradients,
numerical experiments do exhibit preferential inward shifts through random motions of
test particles\,\cite{birmingham1967,walt1971}.
Detailed specification of the fluctuations or the microscopic motion of particles is not 
the subject of present effort.
We plan to construct, instead, a clear-cut description of 
equilibria that maximize entropy simultaneously with bearing
steep density gradients.
Such an equilibrium will be formulated as a grand canonical distribution on
a leaf of foliated phase space that represents a macro-hierarchy.
In a strongly inhomogeneous magnetic field (typically a dipole magnetic field), 
the phase-space metric of \emph{magnetized particles} is distorted; 
thus the projection of the equipartition distribution
onto the flat space of the laboratory frame
yields peaked profile because of the connecting inhomogeneous Jacobian weight.

\section{General Framework}
\label{sec:general_framework}

\subsection{Preliminaries: Poisson algebra and Casimir invariants}
\label{subsec:Possion_Manifold}
A general Hamiltonian system is endowed with a Poisson bracket $\{ a, b\}$
satisfying antisymmetry $\{a,b\}=-\{b,a\}$,
Jacobi's identity $\{\{a,b\},c\} + \{\{b,c\},a\} + \{\{c,a\},b\} =0$,
and Leibniz' property $\{ab, c\} = a\{b,c\} + b\{a,c\}$.
Let $\bm{z}\in X=\mathbb{R}^n$ denote a state vector 
(here we assume that the phase space $X$ is an $n$-dimensional Euclidean space),
and $\partial_{\bm{z}}$ the gradient in $X$.
An observable is a real function on $X$.
We may represent a Poisson bracket as
\begin{equation}
\{ a, b \} = \langle \partial_{\bm{z}} a , \mathcal{J} \partial_{\bm{z}}b \rangle,
\label{Poisson_bracket_general}
\end{equation}
where $\langle \bm{u}, \bm{v}\rangle$ is the inner product of $X$,
and $\mathcal{J}$ (Poisson operator) is an antisymmetric $n\times n$ matrix
(then the antisymmetry and Leibniz' property are satisfied, while Jacobi's identity is conditional\,\cite{morrison98}).
Given a Hamiltonian $H$,
the evolution of an observable $f(\bm{z})$ is described by
\begin{equation}
\frac{{\rmd}}{{\rmd} t} f = \{ f, H \}.
\label{Liouville_general}
\end{equation}

In a \emph{canonical} Hamiltonian system, the Poisson operator is a symplectic matrix;
writing the state vector as $\bm{z}=(q^1, p^1, \cdots, q^m, p^m)$,
\begin{equation}
\mathcal{J}_{\mathrm{c}} :=\left( \begin{array}{ccc}
J_{\mathrm{c}} & 0   & 0 \\
0   & \ddots& 0 \\
0   & 0   & J_{\mathrm{c}}  \end{array} \right),
\quad 
J_{\mathrm{c}} := \left( \begin{array}{cc}
0 & 1 \\
-1& 0    \end{array} \right).
\label{symplectic-canonical}
\end{equation}
Because $\mathrm{Ker}(\mathcal{J}_{\mathrm{c}}) = \{0\}$,
the equilibrium point is given by $\partial_{\bm{z}}H(\bm{z})=0$.
As we see in many examples of so-called weakly coupled systems, 
Hamiltonians are rather simple
---they are often norms of the phase space---
thus the equilibrium points are at most trivial 
(remember the example of a harmonic oscillator).

A general Hamiltonian system may allow the Poisson operator $\mathcal{J}$ to be nontrivial;
it may be a function of $\bm{z}$, and moreover, may have a nontrivial kernel
$\mathrm{Ker}(\mathcal{J}) = \{ \bm{u}\in X;\, \mathcal{J}\bm{u}=0 \}$.
A nontrivial kernel introduces an essential \emph{noncanonicality} to the system, and
brings about interesting structures (Sec.\,\ref{subsec:energy-Casimir}).
If 
\begin{equation}
\partial_{\bm{z}} C \in \mathrm{Ker}(\mathcal{J}),
\label{Casimir-definition}
\end{equation}
such $C(\bm{z})$ is called a \emph{Casimir invariant} (or a \emph{center} of the Poisson algebra).
Evidently, $\{C, G\}=0$ for every $G(\bm{z})$.  Hence, by (\ref{Liouville_general}),
${\rmd} C/{\rmd} t=0$ , i.e. $C(\bm{z})$ is a constant of motion.

Notice that the constancy of $C(\bm{z})$ is independent of the choice of Hamiltonian,
a clear contrast to the more usual invariant that is related to a symmetry of a Hamiltonian.
In later discussion, however, we will connect a Casimir invariant to an \emph{adiabatic invariant},
and then the constancy of a Casimir invariant will be interpreted as a result of
a micro-scale (coarse-grained) symmetry of a Hamiltonian.


\begin{remark}
\label{remark:Lie-Darboux}
Obviously, if $\mathrm{Rank}\,\mathcal{J}(\bm{z})=n$ (the dimension of the phase space), 
(\ref{Casimir-definition}) has only a trivial solution ($C=$ constant).
If the dimension $\nu$ of $\mathrm{Ker}(\mathcal{J}(\bm{z}))$ does not change, 
the solution of (\ref{Casimir-definition}) may be
constructed by ``integrating'' the elements of $\mathrm{Ker}(\mathcal{J}(\bm{z}))$
---then the Casimir leaves are symplectic manifolds. 
This expectation turns out to be true as far as the Poisson bracket satisfies
Jacobi's identity and $n-\nu$ is an even number (Lie-Darboux theorem).
However, the point where $\mathrm{Rank}\,\mathcal{J}(\bm{z})$ changes is the singularity of 
PDE (\ref{Casimir-definition}), 
from which singular Casimir elements are generated\,\cite{yoshida-dewar2012,YMD2014,yoshida_IUTAM2014}.
\end{remark}

\subsection{Energy-Casimir function} 
\label{subsec:energy-Casimir}

When we have a Casimir invariant $C(\bm{z})$ in a noncanonical Hamiltonian system,
a transformation of the Hamiltonian $H(\bm{z})$ such as
(with an arbitrary real constant $\mu$)
\begin{equation}
H(\bm{z}) \mapsto {H}_{{\mu}}(\bm{z})
= {H}(\bm{z}) - \mu {C} (\bm{z})
\label{Hamiltonian-system-3}
\end{equation}
does not change the dynamics.
In fact, the equation of motion (\ref{Liouville_general}) is invariant under this transformation.
We call the transformed Hamiltonian 
${H}_{{\mu}}(\bm{z})$ an \emph{energy-Casimir} function\,\cite{morrison98}.

Interpreting the parameter $\mu$ as a Lagrange multiplier of variational principle,
${H}_{{\mu}}(\bm{z})$ is the effective Hamiltonian with the constraint restricting the
Casimir element $C(\bm{z})$
to have a fixed value (since $C(\bm{z})$ is a constant of motion, its value is fixed at the
initial value).
Even when a Hamiltonian is simple,
an energy-Casimir functional may have a nontrivial structure.
Geometrically, ${H}_{{\mu}}(\bm{z})$ is the distribution of $H(\bm{z})$ on a 
Casimir leaf (a surface of $C(\bm{z})=$ constant).  If Casimir leaves are
distorted with respect to the energy norm, the effective Hamiltonian ${H}_{{\mu}}(\bm{z})$
may have complex distribution on the leaf, which is, in fact, the origin of
various interesting structures in noncanonical Hamiltonian systems.

\subsection{Grand canonical ensemble} 
\label{subsec:grand-canonical}

The foliated phase space of a noncanonical Hamiltonian system can be
viewed as an ensemble of a constrained system
---a Casimir invariant, representing the constraint,
is often regarded as a ``charge'' of the system
(the conservation of charge is an \emph{a priori} condition of dynamics, which is independent of the Hamiltonian).
To formulate the statistical mechanics for such a system, 
we consider a \emph{grand canonical ensemble} determined by a total charge $M$,
in addition to the standard determinant, the total particle number $N$ and the
total energy $E$.
The equilibrium is, then, the maximizer of the entropy $S= - \int f \log f {\rmd}^n z$
under the constraints on the particle number $N= \int f {\rmd}^n z$, 
the energy $E = \int H f {\rmd}^n z$, and the charge $M=\int C f {\rmd}^n z$;
the variational principle
\begin{equation}
\delta ( S - \alpha N - \beta E - \gamma M) =0
\label{variationa}
\end{equation}
yields a Boltzmann distribution
\begin{equation}
f(\bm{z}) = Z^{-1} {\rme}^{-\beta H-\gamma C},
\label{Boltzmann-general}
\end{equation}
where $Z$ ($={\rme}^{\alpha+1}$) is the normalization factor,
$\beta$ is the inverse temperature, and $\gamma/\beta$ is the
\emph{chemical potential} measuring the energy brought about by a change in the charge.

\begin{remark}
\label{remark-Nambu}
One may interpret (\ref{Boltzmann-general}) 
as a Boltzmann distribution with two different energies $H$ and $C$ (with the
corresponding inverse temperatures $\beta$ and $\gamma$).
Here, we remind the pioneering work of Nambu\,\cite{Nambu},
in which a similar grand canonical distribution function was derived for
a ``generalized Hamiltonian system'' with two Hamiltonians on a 
$\mathrm{SO(3)}$ configuration space
---the second Hamiltonian corresponds to a Casimir invariant in the present framework.
\end{remark}

\subsection{Diffusion on distorted phase space}

How a density $f$ depends on the metric of the phase space ($n$ dimensional) 
is formulated by identifying it as a differential $n$-form (or, an $n$-covector).
It is essential to distinguish an extensive quantity $f$ and an intensive quantity $\phi$;
the former (latter) is an $n$-form (a 0-form);
the former transforms as [with a Jacobian weight $\mathrm{D}(y^1,\cdots,y^n)/\mathrm{D}(x^1,\cdots,x^n)$]
\[
f(x^1,\cdots,x^n) = f(y^1,\cdots,y^n)\frac{\mathrm{D}(y^1,\cdots,y^n)}{\mathrm{D}(x^1,\cdots,x^n)},
\]
while the latter is independent of the coordinate transformation.
This is because $f(x^1,\cdots,x^n){\rmd} x^1\wedge\cdots\wedge{\rmd} x^n$
(instead of $f$ alone) represents 
a physical number, and is exactly at the core of the calculus performed in the foregoing subsections.

In the theoretical foundation of statistical mechanics,
the \emph{invariant measure} based on Liouville's theorem (corresponding to the
Poisson bracket of the system) is of fundamental significance.
A diffusion equation (or, a collision operator), therefore, must be formulated 
in consistency with the invariant measure. The equilibrium state is, then,
given by maximizing entropy with respect to the invariant measure. 
Given an invariant measure
${\rmd} x^1\wedge\cdots\wedge{\rmd} x^n$, the diffusion equation governing $f$ and $\phi$ are,
respectively,
\begin{eqnarray}
\partial_t f &=& {\rmd} (\mathcal{D} \delta f),
\label{diffusion-n}
\\
\partial_t \phi &=& \delta(\mathcal{D} {\rmd}\phi).
\label{diffusion-0}
\end{eqnarray}
where $\mathcal{D}$ is a diffusion coefficient,
${\rmd}$ is the exterior derivative (gradient) and $\delta := (-1)^{n+1}~* {\rmd}*$ is the codifferential
($*$ is the Hodge star operator). 
Thus, the diffusion of $f$ is a process of flattening $*f = f * {\rmd} x^1\wedge\cdots\wedge{\rmd} x^n$, while that of $\phi$ is simply the flattening of $\phi$.
When we observe the diffusion on some reference frame with coordinates $y^1,\cdots,y^n$,
the volume element ${\rmd} x^1\wedge\cdots\wedge{\rmd} x^n$ may be inhomogeneous,
and then, the diffusion results in creating an inhomogeneous density $f(y^1,\cdots,y^n)$.
In the next section, we will see such an example of \emph{distorted metric} 
caused by the inhomogeneity in the magnetic field;
instead of the conventional Lebesgue measure of the flat Galilean space,
the flux-tube volume is invariant, and this is the root cause of the inward diffusion observed in magnetospheres
(for example, \cite{Siscoe} formulates a diffusion equation for the flux-tube density of a magnetospheric plasma).

The diffusion of $f$ is caused by fluctuations that violate
conservation of microscopic data (initial conditions of each particle)
while conserving the macroscopic invariants that serve as
the determinants of a statistical ensemble.
What is highly nontrivial is that the diffusion (occurring ``inward'' as
demonstrated, for example, in an electron plasma\,\cite{yoshida2010,saito2010})
is a process creating an inhomogeneous structure.
In a flat (homogeneous metric) space, the equilibrium state is just trivially stable,
while the equilibrium associated with a distorted (inhomogeneous) metric remains stable because
the \emph{free energy} is constrained by the macroscopic constants.
The free energy of a grand canonical system (the logarithm of the grand canonical
partition function) is the sum of the internal energy $E$ and the coupled 
``external'' energies; the latter are measured by ``particle numbers'' multiplied by chemical potentials.

\section{Foliation by adiabatic invariants}

In the foregoing argument, a Casimir invariant was considered as an abstract
constraint on a Hamiltonian system;
whereas we called $M=\int C {\rmd}^n z$ a total charge of a grand canonical ensemble,
the physical meaning of such a charge has not been identified.
In this section, we study a concrete example in which a Casimir invariant is
equivalent to an adiabatic invariant.
The physical meanings of the foliated phase space and the
Boltzmann distribution on it then become clear.

\subsection{Hamiltonian of charged particle}

As an example of Hamiltonian system that has a hierarchical structure in terms of
adiabatic invariants\,\cite{adiabatic_hierarchy}, 
we study a plasma confined by a magnetic field,
and by which we relate Casimir invariants to adiabatic invariants.

\subsubsection{Magnetic coordinates} 
Here we consider an axisymmetric system with a poloidal (but no toroidal) magnetic field
that can be written as
\begin{equation}
\bm{B}=\nabla\psi\times\nabla\theta,
\label{magnetic_field}
\end{equation}
where $\theta$ is the toroidal angle and $\psi$ is the magnetic flux function
(the Gauss potential of $\bm{B}$).
Let $\zeta$ be the parallel coordinate along each magnetic surface (the
level-set of $\psi$).
We can choose $(\psi,\zeta,\theta)$ as the coordinates of the configuration space
($\theta$ is ignorable in an axisymmetric system).
For example, a point-dipole magnetic field  is represented by  
\begin{equation}
\left\{ \begin{array}{l}
\psi(r,z) = M r^2(r^2+z^2)^{-3/2},
\\
\zeta(r,z) = M z(r^2+z^2)^{-3/2},
\end{array} \right.
\label{point-dipole}
\end{equation}
where $(r,z,\theta)$ are the cylindrical coordinates and $M$ is the magnetic moment.

\subsubsection{Hierarchy of adiabatic invariants} 
The magnetized particles have three different adiabatic invariants, 
i.e., the magnetic moment $\mu$, the action $J_\parallel$ of bounce motion, 
and the action (canonical angular momentum) $P_\theta$ of the toroidal drift\,\cite{adiabatic_hierarchy}.
When the magnetic field is sufficiently strong, the corresponding frequencies define a hierarchy:
$\omega_c$ (cyclotron frequency) $\gg$ $\omega_b$ (bounce frequency) $\gg$ $\omega_d$ (drift frequency).
Hence, $\mu$ is the most robust adiabatic invariant.  
On the other hand, the constancy of $P_\theta$ is easily broken by a 
large-scale ($\sim$ system size), slow ($\lesssim \omega_d$) perturbations 
destroying the azimuthal symmetry.
In a quasi-neutral plasma ($\phi=0$), we may estimate  $|\bm{v}_d|/|\bm{v}_c| \sim \rho_c/L \ll 1$
($\bm{v}_c$ is the gyration velocity, $\bm{v}_d$ is the toroidal drift velocity,
$\rho_c$ is the gyro-radius, and $L$ is the macroscopic system size).
Neglecting $v_d$ in $P_\theta=mr v_d + q\psi$, we may approximate $P_\theta=q\psi$.

\subsubsection{Hamiltonian} 
The Hamiltonian of a charged particle is the sum of the kinetic energy and the potential energy:
\begin{equation}
H = \frac{m}{2} v^2 + q \phi,
\label{Hamiltonian-1}
\end{equation}
where $\bm{v} := (\bm{P}-q\bm{A})/m$ is the velocity,
$\bm{P}$ is the canonical momentum, $(\phi,\bm{A})$ is the electromagnetic 4-potential,
$m$ is the particle mass, and $q$ is the charge.
In the present work, we may treat electrons and ions equally.
In a non-neutral plasma, $\phi$ includes the self-electric field that plays an essential role in
determining the equilibrium\,\cite{yoshida2010,non-neutral}.

In order take into account the aforementioned hierarchy of actions,
we invoke a canonical phase space of action-angle pairs;
(denoting the gyro angle by $\vartheta_c$ and the bounce angle by $\vartheta_b$)
\begin{equation}
\bm{z} = (\mu,\vartheta_c; J_\parallel,\vartheta_b; \psi,\theta),
\label{canonical_coordinates}
\end{equation}
and write the Hamiltonian of a particle as
\begin{equation}
H_{gc} = \omega_c  \mu + \omega_b J_\parallel + q \phi.
\label{GCHamiltonian}
\end{equation}
Here, we have omitted the kinetic energy of the toroidal drift velocity by approximating $P_\theta=q\psi$\,\cite{drift-kinetic}.
The gyro angle is coarse grained (averaged out), so it is eliminated in $H_{gc}$
(i.e., $H_{gc}$ dictates the motion of the guiding center of the gyrating particle).
In the standard interpretation, in analogy with the Landau levels in quantum theory,
$\omega_c$ is the \emph{energy level}
and $\mu$ is the \emph{number} of quasi-particles (quantized guiding center) 
at the corresponding energy level\,\cite{hbar};
the term $\omega_c  \mu$ in $H_{gc}$ represents the macroscopic (classical) energy of the
quasi-particles.

\subsection{Foliation by adiabatic invariants}
\label{subsec:foliation_by_ad_inv}

\subsubsection{Foliation by $\mu$} 
To extract the macro-hierarchy, we separate
the microscopic variables $(\vartheta_c,\mu)$ 
by modifying the Poisson matrix as
\begin{equation}
\mathcal{J}_{\mu} :=\left( \begin{array}{ccc}
0 & 0   & 0 \\
0 & J_{\mathrm{c}} & 0 \\
0 & 0   & J_{\mathrm{c}} \end{array} \right).
\label{symplectic-noncanonical}
\end{equation}
The Poisson bracket
\[
\{ F, G \}_{\mu} := \langle\partial_{\bm{z}}F, \mathcal{J}_{\mu}\partial_{\bm{z}}G\rangle
\]
determines the kinematics on the macro-hierarchy that separates the canonical pair $(\mu,\vartheta_c)$.

The nullity of $\mathcal{J}_{\mu}$ makes the Poisson bracket $\{\, ,\,\}_{\mu}$ \emph{noncanonical}.
Evidently, $\mu$ is a Casimir invariant
(more generally every $C=g(\mu)$ with $g$ being any smooth function is a Casimir invariant).
The level-set of $\mu$, a leaf of the Casimir foliation, identifies
what we may call the \emph{macro-hierarchy}.

\begin{remark}
The adiabatic invariant $\mu$ appears in several manifestations; it has been called
 ``Casimir invariant'', ``charge'', and ``particle number''. Although these names are used synonymously,
their specific conservations carry different implications.
As noted above, the Landau level analogy  allows us to assign an adiabatic invariant with a particle number,
and this interpretation plays an essential role in  formulating the ``grand canonical ensemble'' 
(Sec.\,\ref{subsec:grand-canonical}) in which
 $\mu$ plays the role of a quasi-particle number (Sec.\,\ref{subsec:Boltzmann}). 
The equivalent christening  of $\mu$ as  ``charge'' or a Casimir invariant is rather profoundly motivated.
In a noncanonical Hamiltonian system, the nullity of the Poisson bracket
yields a \emph{topological charge}, i.e., a Casimir invariant (Sec.\,\ref{subsec:Possion_Manifold}).
Such a topological charge can be related to a \emph{Noether charge}
pertinent to a symmetry of some appropriate action principle.
When a Casimir invariant is an adiabatic invariant (as constructed here), 
it is a consequence of symmetry with respect to a coarse-grained angle $\vartheta$. 
The following mathematical formality provides content to the preceding  statement. 
Let $S = \int \Lambda - H \rmd t$ be a microscopic (canonical) action,
where $\Lambda$ is a canonical 1-form and $H$ is a Hamiltonian.
Suppose that $\vartheta$ is a microscopic angle of some periodic motion,
which gives an adiabatic invariant $J=\oint P_\vartheta \rmd\vartheta/2\pi$.
Then a macroscopic action $S'$ can be defined by separating, from $S$,
a microscopic action $S_J = \int (J \dot{\vartheta} - \omega J)\rmd t$
($\omega J$ is the energy of the microscopic periodic motion with frequency $\omega$).
Averaging over the periodic motion, $S'$ is made to be independent of $\vartheta$.
The symmetry $\partial_\vartheta =0$ of the coarse-grained action $\overline{S}=S' + S_J$ yields a
Noether charge $\partial_{\dot{\vartheta}}\overline{S}=J$,
i.e., the Casimir invariant
(the variation of the macroscopic part $S'=\int \Lambda' - H'\rmd t$
yields a degenerate 2-form $\rmd \Lambda'$ whose nullity is spanned by the Casimir invariant).
\end{remark}

\subsubsection{Foliation by $J_\parallel$} 
We may define a more macroscopic hierarchy by separating the
second canonical pair $(\vartheta_b,J_\parallel)$ from the phase space.
In comparison with the previous process of defining $\{~,~\}_\mu$,
we need somewhat complicated procedure, because 
the bounce angle ($\vartheta_b$) is not ignored in $H_{gc}$;
the frequencies $\omega_c$ and $\omega_b$ (as well as $\phi$ in a non-neutral plasma) 
are functions of the spacial coordinates including $\vartheta_b$
($\zeta=\ell_\parallel\sin\vartheta_b$ with the bounce orbit length  $\ell_\parallel$).
We have
\begin{equation}
\dot{J}_\parallel=\frac{\partial H_{gc}}{\partial \vartheta_b} .
\label{bounce_action}
\end{equation}
For the periodic bounce motion, 
$\oint (\partial H_{gc}/\partial \vartheta_b)d\vartheta_b =\oint d H_{gc}=0$.
Integrating (\ref{bounce_action}) over the cycle of bounce motion yields
the bounce-average $\langle J_\parallel\rangle=$ constant.
When we calculate macroscopic quantities (like the total energy or the total action),
we evaluate $J_\parallel$ as the adiabatic invariant $\langle J_\parallel\rangle$,
and then the second action-angle pair $(\vartheta_b, J_\parallel)$ is separated from the dynamical variables;
the corresponding Poisson matrix is
\begin{equation}
\mathcal{J}_{\mu J_\parallel} :=\left( \begin{array}{ccc}
0 & 0   & 0 \\
0 & 0   & 0 \\
0 & 0   & J_{\mathrm{c}} \end{array} \right),
\label{symplectic-noncanonical-2}
\end{equation}
and the Poisson bracket is
\[
\{ F, G \}_{\mu J_\parallel} := \langle\partial_{\bm{z}}F, \mathcal{J}_{\mu J_\parallel}\partial_{\bm{z}}G\rangle .
\]

Now, the dynamical variables are only $\theta$ and $\psi$.
The drift frequency is given by bounce-averaging the toroidal angular velocity
\begin{equation}
\omega_d=\dot{\theta} = \frac{\partial H_{gc}}{\partial\psi} 
=\mu \frac {\partial\omega_c }{\partial \psi}+ J_\parallel\frac{\partial\omega_b }{\partial \psi} 
 + q \frac{\partial \phi}{\partial\psi} ,
\label{drift_frequency}
\end{equation}
As long as the system maintains the toroidal symmetry $\partial/\partial\theta=0$,
the third action $\psi$ remains constant, and the orbit of the guiding center is completely integrable.
A slow perturbation, however, may break the constancy of $\psi$, allowing the guiding center to
cross magnetic surfaces.
As shown in the next section,
the Boltzmann distribution on this minimum (most macroscopic) phase space
has an interesting structure.

\subsection{Boltzmann distributions}
\label{subsec:Boltzmann}

\subsubsection{Microscopic phase space} 
The standard Boltzmann distribution function is derived when we assume that 
${\rmd}^6 z= {\rmd}^3v {\rmd}^3x$ is an invariant measure and the
Hamiltonian $H$ is the determinant of the ensemble. 
Maximizing the entropy $S$
keeping the total energy $E$
and the total particle number $N$
constant, we obtain
\begin{equation}
f(\bm{x},\bm{v}) = Z^{-1} {\rme}^{-\beta H}.
\label{Boltzmann-1}
\end{equation}
The corresponding configuration-space density is
\begin{equation}
\rho(\bm{x}) = \int f {\rmd}^3 v \propto {\rme}^{-\beta q\phi},
\label{density-1}
\end{equation} 
which becomes constant for an electrically neutral system ($\phi=0$).

Needless to say, the Boltzmann distribution or the corresponding configuration-space density,
with an appropriate Jacobian multiplication, is independent 
of the choice of phase-space coordinates.
Moreover, the density is invariant 
no matter whether we coarse grain the cyclotron motion or not.
Let us confirm this fact by a direct calculation.
The Boltzmann distribution of the \emph{guiding-center plasma} is
\begin{eqnarray}
f(\bm{z}) &=& Z^{-1} {\rme}^{-\beta H_{gc}} 
\nonumber \\
&=&
Z^{-1} {\rme}^{-\beta\left[ m(v_\perp^2 + v_\parallel^2)/2  + q \phi \right]},
\label{Boltzmann-2}
\end{eqnarray}
where $\bm{v}_\perp$ and $\bm{v}_\parallel$ are the perpendicular and parallel 
(with respect to the local magnetic field) components of the velocity.
Here we neglected the kinetic energy of the drift motion to approximate $\omega_c\mu\approx mv_\perp^2/2$.
The corresponding density reproduces (\ref{density-1}).

\subsubsection{Boltzmann distribution on the $\mu$ leaf} 
Now we calculate the Boltzmann distribution on the \emph{macro-hierarchy}.
We start with the Casimir leaf of $\mu$.
The adiabatic invariance of  $\mu$
imposes a topological constraint on the motion of particles;
this constraint is the root cause of a macro-hierarchy and of structure formation. 
By applying Liouville's theorem to the Poisson bracket $\{\, ,\,\}_{\mu}$, 
the invariant measure on the macro-hierarchy (the $\Gamma$-space of
quasi-particles) is 
\[
\prod_j {\rmd}^4z_j= \prod_j {\rmd}^6z_j/(2\pi {\rmd}\mu_j),
\]
i.e., the total phase-space measure modulo the
microscopic measure (suffix $j$ is the index of each particle).
The most probable state (statistical equilibrium) on the macroscopic ensemble must
maximize the entropy with respect to this invariant measure.

To determine the distribution function, the variational principle is set up 
immersing the ensemble into the general phase space, 
and incorporating the constraints through the Lagrange multipliers:
We maximize entropy $S=-\int f\log f\,{\rmd}^6z$ 
for a given particle number $N= \int f {\rmd}^6z$, a quasi-particle number $M_1=\int \mu f {\rmd}^6z$,
and an energy $E=\int H_{gc} f {\rmd}^6z$,
to obtain the distribution function (see Sec.\,\ref{subsec:grand-canonical})
\begin{equation}
f = f_\gamma := Z^{-1} {\rme}^{ -(\beta H_{gc} + \gamma\mu)}.
\label{modified_Boltzmann}
\end{equation}

\begin{figure}[tb]
\label{fig:density}
\begin{center}
\includegraphics[scale=0.6]{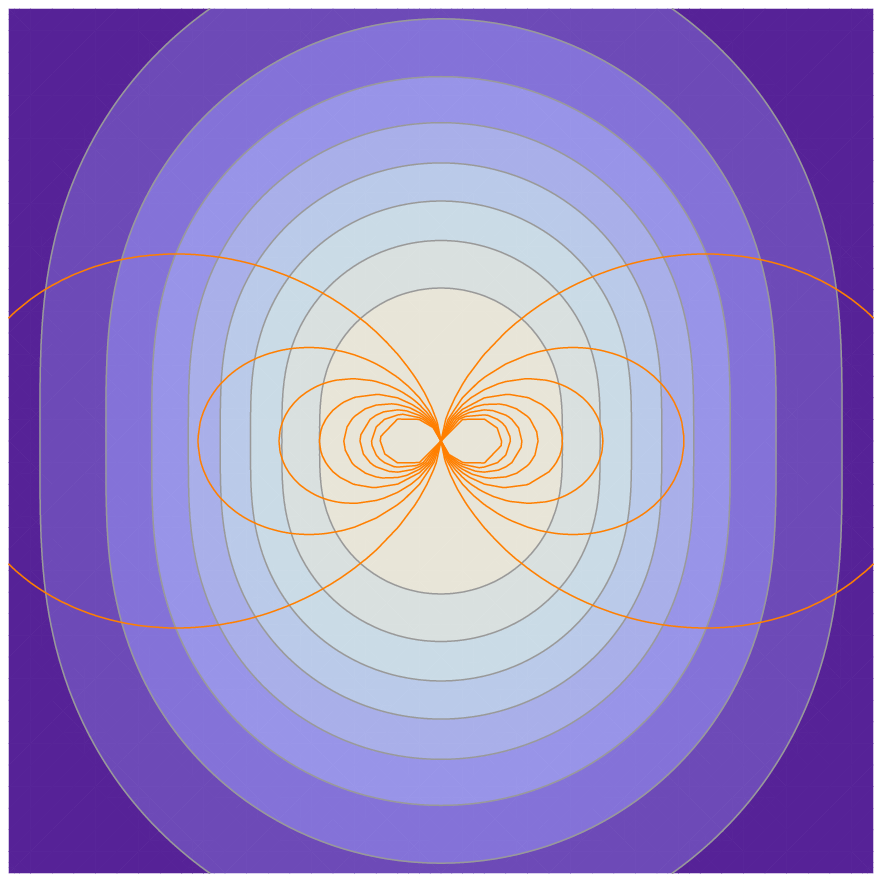}
~~
\includegraphics[scale=0.6]{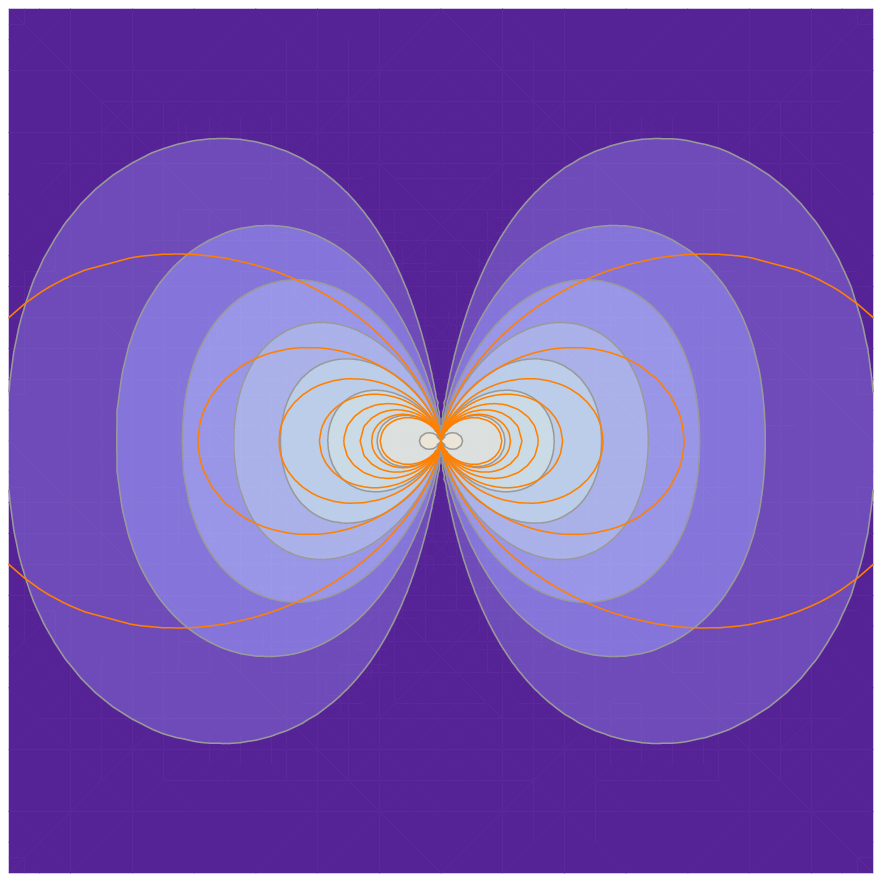}
\caption{
Density distribution (contours) and the magnetic field lines (level-sets of $\psi$)
in the neighborhood of a point dipole. Left: The equilibrium on the
leaf of $\mu$-foliation. Right: The equilibrium on the
leaf of $\mu$ and $J_\parallel$-foliation. 
}
\end{center}
\end{figure}

The factor ${\rme}^{-\gamma\mu}$ in $f_\gamma$
yields a direct $\omega_c$ dependence of the coordinate-space density:
\begin{equation}
\rho =\int f_\gamma \, \frac{2\pi\omega_c }{m} {\rmd}\mu {\rmd} v_\parallel
\propto \frac{\omega_c(\bm{x})}{\beta\omega_c(\bm{x})+\gamma } .
\label{density-3}
\end{equation}
Here we are assuming electric neutrality to put $\phi=0$.
Notice that the Jacobian $(2\pi\omega_c /m){\rmd}\mu$ multiplying the macroscopic 
measure ${\rmd}^4z$ reflects the distortion of the macroscopic phase space (Casimir leaf)
caused by the magnetic field.
Figure\,\ref{fig:density}-(left) shows the density distribution and the magnetic field lines.

\subsubsection{Boltzmann distribution on the $\mu$-$J_\parallel$ leaf} 
We may further restrict the second action $J_\parallel$ and
calculate the Boltzmann distribution of the $\mu$-$J_\parallel$ leaf.
Imposing another constraint on $M_2=\int J_\parallel f {\rmd}^6z$,
we modify (\ref{modified_Boltzmann}) as
\begin{equation}
f_{\gamma_1,\gamma_2} = Z^{-1} {\rme}^{ -(\beta H_{gc} + \gamma_1\mu+\gamma_2 J_\parallel)}.
\label{modified_Boltzmann2}
\end{equation}
To find explicit expressions for the parallel action-angle variables,
let us solve the equation of parallel motion under some approximations.
Neglecting the curvature of magnetic field lines and
putting $\phi=0$, 
\begin{equation}
m \ddot{\zeta} = -\mu \nabla_\parallel{\omega_c} ,
\label{parallel_motion2}
\end{equation}
where $\nabla_\parallel := \bm{b}\cdot\nabla$ with the magnetic unit vector $\bm{b}:=\bm{B}/B$.
In the vicinity of $\zeta=0$, where $\omega_c$ has a minimum on each magnetic surface, we may approximate
\[
\omega_c= \Omega_{c}(\psi) + \Omega_c''(\psi) \frac{\zeta^2}{2} ,
\]
where $\Omega_{c}(\psi)$ is the minimum of $\omega_c$ on each contour of  $\psi$,  
and $\Omega_c''(\psi) := {\rmd}^2\omega_c/{\rmd}\zeta^2|_\psi$.
Integrating  (\ref{parallel_motion2}), we obtain a harmonic oscillation 
with the
bounce frequency
\begin{equation}
\omega_b = \sqrt{\frac{\Omega_{c}''(\psi)\mu}{m}}
= \frac{v_\perp }{L_\parallel(\psi)},
\label{bounce-frequency}
\end{equation}
where $L_\parallel(\psi):=\sqrt{2\Omega_c(\psi)/\Omega_c''(\psi)}$ is the length scale of 
the variation of $\omega_c$ along $\zeta$.
The amplitude of the oscillation, i.e., the bounce orbit length is
\begin{equation}
\ell_\parallel = \sqrt{\frac{2E_\parallel}{m\omega_b^2}},
\label{orbit_length}
\end{equation}
where  $E_\parallel:=(mv_\parallel^2)/2|_{\zeta=0}$ is the kinetic energy of the parallel motion. 
Assuming $E_\parallel\approx E_\perp:=\mu\Omega_c$, we may estimate
$\ell_\parallel\approx L_\parallel$.
By $E_\parallel=\omega_b J_\parallel$, we obtain
\[
{\rmd}v_\parallel=\frac{\omega_b}{mv_\parallel}{\rmd}J_\parallel
=\sqrt{\frac{\omega_b}{2mJ_\parallel}}{\rmd}J_\parallel .
\]
Using the relation $\omega_b/(mv_\parallel)=v_\perp/(L_\parallel m v_\parallel)
\approx 1/(mL_\parallel)$, we may write 
\begin{equation}
{\rmd}v_\parallel \approx \frac{{\rmd}J_\parallel}{mL_\parallel}.
\label{dJ_parallel}
\end{equation}
The density is given by
\begin{eqnarray}
\rho &=& \int f_{\gamma_1,\gamma_2} \, \frac{2\pi\omega_c {\rmd}\mu}{m} \frac{{\rmd} J_\parallel}{m L_\parallel} .
\nonumber \\
&\propto& \frac{\omega_c(\bm{x})}{m^2}\int^\infty_0 \frac{{\rme}^{-(\beta\omega_c+\gamma_1)\mu} {\rmd}\mu}{\beta\sqrt{2\omega_c\mu/m}+\gamma L_\parallel(\psi)}.
\label{density-4}
\end{eqnarray}
Numerical integration of (\ref{density-4})
gives the density profile depicted in Fig.\,\ref{fig:density}-(right)\,\cite{YoshidaPPCF2013}.

\begin{remark}
\label{remark:grand-canonical}
The derived distribution function $f_{\gamma_1,\gamma_2}$
is a particular solution of the stationary kinetic equation $\{H_{gc}, f\}_{\mu J_\parallel}=0$.
This thermodynamic equilibrium, however, has deeper meaning
than the arbitrary solutions such as $f=F(\mu,J_\parallel,\psi)$
that are often invoked in drift-kinetic calculations.
For instance, $f=F(\mu,J_\parallel)$ 
yields a density $\rho\propto \omega_c/L_\parallel$ 
(implying that the particle number per unit flux tube distributes homogeneously), 
which in a dipole magnetic field, scales as $\propto r^{-4}$, precisely the density profile given by
Hasegawa\,\cite{hasegawa2005}
(for example, see \cite{Schippers} for realization of a similar distribution in Saturn's magnetosphere).
Choosing $F$ to be a Gaussian, $Z^{-1} e^{-(\gamma_1\mu+\gamma_2 J_\parallel)}$ is the asymptotic form
of (\ref{density-4}) in the limit $r\rightarrow\infty$ ($\omega_c \rightarrow 0$
so that $\beta H_{gc} \ll \gamma_1 \mu+\gamma_2 J_\parallel$).
Such a solution is also the $\beta\rightarrow0$ (infinite temperature) 
limit of $f_{\gamma_1,\gamma_2}$. For finite temperatures, 
the energy constraint prevents the particle
distributing homogeneously on the ensemble foliated by $\mu$ and $J_\parallel$.
Notice that for the distribution $f_{\gamma_1,\gamma_2}$, the density $\rho$ remains finite, while for the  solution $f=F(\mu,J_\parallel)$, it diverges as  $\omega_c \rightarrow \infty$. 
For experimental evidence of density limitation, see \cite{saito2011}.
\end{remark}

\begin{remark}
\label{remark-self-field}
In the foregoing analysis, we did not pay attention to the field equation (Maxwell's equation),
and dealt with the magnetic flux function $\psi$ as a given function of space.
However, when the plasma pressure becomes comparable to the pressure of the dipole magnetic field
(i.e., the so-called beta ratio is of order unity; see\,\cite{YoshidaPPCF2013}), 
we have to adjust the magnetic field to take into account the spontaneous component.
This can be done by solving the Grad-Shafranov equation for $\psi$ 
with the plasma pressure given by the distribution function
(to take into account the pressure anisotropy, we have to use the generalized Grad-Shafranov equation\,\cite{Grad1967,Furukawa2014}).
We also assumed charge neutrality, and put the electric potential $\phi=0$.
In a non-neutral plasma\,\cite{yoshida2010}, the spontaneous electric potential $\phi$ must be
determined by solving the Poisson equation
(the grand canonical ensemble of a non-neutral plasma is also constrained by the third adiabatic invariant $P_\theta \approx q\psi$,
because the total angular momentum plays an important role in ``neutralizing'' the spontaneous electric field on the comoving frame; see also \cite{Gubar}).
The field equations (the Grad-Shafranov equation for $\psi$, and the Poisson equation for $\phi$)
pose a nonlinear problem, because we have to find a self-consistent distribution function that depends on $\psi$ and $\phi$.
Numerical analyses of these equations will be presented elsewhere.
\end{remark}

\section{Summary and concluding remarks}

In this paper, we have developed a conceptual framework for delineating and understanding
the advanced notion of self-organization simultaneous with entropy production. 
An appropriate \emph{scale hierarchy}, encompassing large-scale order and small-sale disorder, is established 
by exploiting phase-space foliation provided by the adiabatic invariants of the system;
the corresponding invariant measure is also specified. 
A leaf of the foliated phase space is identified as a grand canonical ensemble of
macroscopic quasi-particles representing coarse grained (averaged over microscopic angle of periodic motion) orbits.

As an explicit example, we have constructed a foliated phase space representing the scale hierarchy of
magnetized particles in a magnetospheric plasma.
The Boltzmann distribution is obtained by maximizing the entropy for a given particle number and a quasi-particle number as well as a total energy.  
The system is driven to such a Boltzmann distribution by some entropy production mechanism that, inherently, preserves the adiabatic invariants (Casimir invariants). 
The spatiotemporal scales of associated fluctuations must be larger than the scales on which the  conjugate coarse-grained angle variables vary. 
Under the same condition on possible perturbations, 
the Boltzmann distribution on the macro-hierarchy is absolutely stable, 
because it is the minimizer of the energy (as an isolated system).
It is interesting that the steep density gradient predicted by the distribution (\ref{density-4})  is stable against 
macroscopic modes such as interchange modes (cf.\,\cite{Siscoe} );
an intuitive explanation is that the magnetized particles reside in the magnetic-coordinate space, where the
actual density distribution is flat, leaving no free energy for instabilities.

The derived grand canonical distribution function opens a
new channel for extracting interesting properties of magnetized plasmas.
For example, $\mu$ could be boosted by cyclotron heating (for magnetically confined particles), and  
the resulting increase in the total magnetic moment
of the system could cause macroscopic motion of the
levitated magnet accompanied by the plasma\,\cite{yoshida2012}.
In our model, an increase of $\mu$ means injection of quasi-particles;
an increased quasi-particle number $M$, in turn, automatically increases the 
macroscopic magnetic moment.
This simple picture is beyond the reach of the conventional canonical (or micro-canonical) distributions
that are unaware of any direct relation between the macroscopic magnetic moment
and $\mu$ [even if we write $mv_c^2/2 = \mu\omega_c$ as in (\ref{Boltzmann-2})].
Heating, therefore, could not create or destroy magnetic moment (or any axial vector);
the coupling of heat and mechanical energy could manifest only through the pressure force.
Our distribution function, on the other hand, is capable of delineating such connections since it embodies magneto-fluid-thermo couplings.
Many other applications such as estimate of fluctuations, phase equilibrium relations, possible condensation at low temperature, etc. will become accessible through the grand canonical distribution.

The framework we have developed will apply to general systems with 
nontrivial \emph{topologies}.
Viewing from a different angle, 
our work has a common perspective with Nambu's ``generalized Hamiltonian system'' 
that has two Hamiltonians (one of which is a Casimir invariant in the present terminology);
see \cite{Nambu} and Remark\,\ref{remark-Nambu}.
In the present theory, connecting a Casimir invariant further to an adiabatic invariant, 
we have written a kinetic-thermodynamic theory with a built in \emph{scale hierarchy}. 
Structure formation is a direct consequence of embedding the Casimir leaf (where the microscopic actions are abstracted as quasi-particle numbers) into the laboratory flat space.

We end this paper with some comments on self-organization in fluid-mechanical systems.
The dynamics in some fluid models can be cast into a unified Hamiltonian form\,\cite{morrison98}
\begin{equation}
\partial_t\omega = \mathcal{J}\partial_\omega H(\omega),
\label{Hamilton-function_space}
\end{equation}
where $\omega$ is a state vector belonging to a Hilbert space $X$,
$H(\omega)$ is the Hamiltonian which is a real-valued functional on $X$, and $\mathcal{J}$ is the Poisson operator.
The Poisson bracket is defined by $\langle \partial_\omega F, \mathcal{J}\partial_\omega G\rangle$,
where $\langle~,~\rangle$ is the inner-product of $X$.
The vortex equation of two-dimensional Eulerian (inviscid, incompressible) flow is the simplest example;
with a Poisson operator $\mathcal{J}=[\omega, \circ ]$
(where $\omega$ is the vorticity, and $[a, b]= \partial_y a \cdot\partial_x b - \partial_x a \cdot\partial_y b$),
and a Hamiltonian $H(\omega)=\int \omega\cdot(-\Delta)^{-1}\omega\,\rmd^2x/2$
(where $\Delta$ is the two-dimensional Laplacian, and $\Delta^{-1}$ is its inverse),
(\ref{Hamilton-function_space}) reads $\partial_t\omega=[\omega,\phi]$,
(where $\phi=(-\Delta)^{-1}\omega$ is the Gauss potential of the flow).
Notice that $\mathcal{J}$ depends on the dynamical variable $\omega$.
Evidently $C_f = \int f(\omega)\,\rmd^2x$ ($f$ is an arbitrary $C^2$-class function) is a Casimir invariant.
Slightly modifying the Poisson operator as ${\mathcal{J}}=[\omega - g,\circ]$ with 
an inhomogeneous term $g$ (cf.\,\cite{Weinstein}), 
and the Hamiltonian as ${H}=\int \omega\cdot L^{-1}\omega\,\rmd^2x/2$ with $L=-\Delta + 1$
(the term $+1$ in the operator $L$ reflects the compressibility of the fluid),
Hamilton's equation (\ref{Hamilton-function_space}) becomes $\partial_t\omega=[\omega-g,\phi]$
with $\phi=L^{-1}\omega$,
which is formally the Hasegawa-Mima equation~\cite{Hasegawa-Mima} of drift waves in a magnetized plasma
($g$ represents the inhomogeneity of the equilibrium plasma density), 
or the Charney equation~\cite{Charney} of Rossby waves
($g$ represents the inhomogeneity of the Coriolis force and the depth of the atmospheric fluid).
The Casimir invariant is $C_f = \int f(\omega-g)\,\rmd^2x$.
\\
\indent
A theory of self-organization can be described by invoking the scenario of \emph{selective dissipation}
which compares
different constants of motion, the Hamiltonian (energy) and some Casimir invariants
(choosing $f(\omega)=\omega^2/2$, $C_f$ is the \emph{enstrophy});
a functional including higher-order spatial derivatives is more fragile in comparison with a lower-order one,
because small-scale turbulence can dissipate it more easily, thus the energy conserves better than the enstrophy.
Minimizing the enstrophy for a fixed energy, we obtain a ``relaxed state''.
The review paper by Hasegawa\,\cite{Hasegawa1985} describes a list of successful applications of this model,
including the creation of zonal flows by Rossby wave turbulence, and the Taylor relaxed state\,\cite{JBT} by magnetohydrodynamic (MHD) turbulence (which dissipates the energy while conserving the magnetic helicity,
a Casimir invariant of MHD system).
\\
\indent
The same minimization principle has a different connotation, which aligns with the present theory.
The target functional to be minimized is nothing but the energy-Casimir functional
(see Sec.\,\ref{subsec:energy-Casimir}), 
thus the minimizer is an equilibrium point of the Hamiltonian mechanics
(by the duality of well-posed variational principles\,\cite{YoshidaMahajan2002},
the minimizer of the enstrophy for a given energy is equivalent to the maximizer of the energy for a given enstrophy, and the latter is also an equilibrium point of the Hamiltonian system).
To put this equilibrium point into the perspective of the entropy principle, we have to consider
statistical mechanics on a function space.
In \cite{ItoYoshia1996}, a grand canonical ensemble of MHD is formulated by considering ``magnetic-helicity quantum.''
The Taylor relaxed state is, then, the \emph{low-temperature limit} (ground state) of the Boltzmann distribution
(here, the \emph{temperature} measures the strength of the turbulence).
\\
\indent
While the present theory of self-organization has wide applications encompassing particle models to fluid models,
the connection between the two formalisms awaits further exploration.
For example, the helicities are the determinants of the
macroscopic hierarchy (Casimir leaf) of the MHD system,
which control the bifurcation of various equilibrium points by shifting the leaf in the phase space\,\cite{yoshida-dewar2012}.
However, we have not yet unearthed the coarse-grained ``angle variable'' corresponding to the magnetic helicity.
If we can specify the origin of the fluid Casimir invariants in the particle model, we will be able to write a kinetic theory of far richer structures that have various helicities (or, vorticities and currents twisting the stream lines and magnetic field lines). 
\\ 
\indent
The foliation is also an interesting subject to be explored in the 
framework of  a new variable --the entropy production rate.
Since, for a driven (or open) system, the standard entropy is no longer an effective state variable to characterize long-lived structures,
the entropy production rate ($\sigma$) has been deemed to be an alternative determinant of the state.
Debates have raged whether the organizing principle is the minimum or maximum of $\sigma$.
While the minimum $\sigma$ principle\,\cite{dissipation}
applies to linear systems,
typical nonlinear fluid-mechanical systems, instead prefer maximum $\sigma$ states.
Sawada\,\cite{Sawada} proposed a nonlinear mechanism that maximizes $\sigma$;
for other models of fluids and plasmas, see.\,\cite{Ozawa,YM2008,Niven}. 
Dewar\,\cite{RoderickDewar} proposed to evaluate the probability of phase-space trajectories (instead of points),
which is shown to have a Boltzmann distribution by replacing the minus energy by the entropy production along each trajectory;
hence the maximum entropy production is most probable
(see also \cite{KawazuraYoshida2010,YoshidaKawazura2014} for the thermodynamic duality of the maximum and minimum principles that switch depending on whether the system is flux driven or force driven).
When the phase space is foliated, such trajectories must be restricted on a Casimir leaf,
and then the structures self-organized by maximizing $\sigma$ will have topological complexity
charged by the Casimir invariants.

\section*{Acknowledgments}

The authors acknowledge the stimulating discussions and suggestions of Professor A. Hasegawa, Professor P. J. Morrison, and Professor R. D. Hazeltine.
ZY's research was supported by the Grant-in-Aid for Scientific Research 
(23224014) from MEXT-Japan.
SMM's research was supported by the US DOE
Grant DE-FG02-04ER54742.



\begin{thebibliography}{99}

\bibitem{morrison98}
P. J. Morrison, 
Rev. Mod. Phys. \textbf{70}, 467 (1998).



\bibitem{schulz1974}
M. Schulz and L. J. Lanzerotti, {\it Particle Diffusion in the Radiation Belts} (Springer, New York, 1974).

\bibitem{brautigam2000}
D. H. Brautigam and J. M. Albert, J. Geophys. Res. {\bf 105}, 291 (2000).

\bibitem{Chen2007}
Y. Chen, G. D. Reeves, and R. H. W. Friedel, Nature Phys. {\bf 3}, 614 (2007).

\bibitem{boxer2010}
A. C. Boxer, R. Bergmann, J. L. Ellsworth, D. T. Garnier, J. Kesner, M. E. Mauel, and P. Woskov,
Nature Phys. {\bf 6}, 207 (2010).

\bibitem{yoshida2010}
Z. Yoshida, H. Saitoh, J. Morikawa, Y. Yano, S. Watanabe, and Y. Ogawa,
Phys. Rev. Lett. {\bf 104}, 235004 (2010).

\bibitem{saito2010}
H. Saitoh, Z. Yoshida, J. Morikawa, Y. Yano, H. Hayashi, T. Mizushima, Y. Kawai, M. Kobayashi, and H. Mikami,
Phys. Plasmas {\bf 17}, 112111 (2010). 

\bibitem{saito2011}
H. Saitoh, Z. Yoshida, J. Morikawa, Y. Yano, T. Mizushima, Y. Ogawa, M. Furukawa, Y. Kawai, K. Harima, Y. Kawazura, Y. Kaneko, K. Tadachi, S. Emoto, M. Kobayashi, T. Sugiura, and G. Vogel,
Nucl. Fusion \textbf{51}, 063034 (2011).

\bibitem{birmingham1967}
T.J. Birmingham, T.G. Northrop, and C.-G. Falthammar, Phys. Fluids \textbf{10}, 2389 (1967).

\bibitem{walt1971}
M. Walt, Space Sci. Rev. \textbf{12}, 446 (1971).

\bibitem{yoshida-dewar2012}
Z. Yoshida and R. L. Dewar,
J. Phys. A: Math. Theor. \textbf{45}, 365502 (2012).

\bibitem{YMD2014}
Z. Yoshida, P. J. Morrison, and F. Dobarro,
J. Math. Fluid Mech. \textbf{16}, 41 (2014).

\bibitem{yoshida_IUTAM2014}
Z. Yoshida,
Procedia IUTAM \textbf{7}, 141 (2013).

\bibitem{Nambu}
Y. Nambu, 
{Phys. Rev.} D \textbf{7}, 2405 (1973).

\bibitem{Siscoe}
G. L. Siscoe and D. Summers,
J. Geophys. Res. Space Phys. \textbf{86}, 8471 (1981).


\bibitem{adiabatic_hierarchy}
A. J. Lichtenberg and M .A. Lieberman, \textit{Regular and Chaotic Dynamics}, 2nd ed. (Springer-Verlag, New York, 1992), Sec.\,2.3b.


\bibitem{non-neutral}
D. H. E. Dubin and T. M. O'Neil, Rev. Mod. Phys. \textbf{71}, 87
(1999).

\bibitem{drift-kinetic} 
There is a subtlety in the representation of the energy of the drift velocity,
if it were retained in the Hamiltonian formalism in order to analyze the dynamics 
(in response to the electromagnetic perturbations) of the guiding center; 
see J. R. Cary and A. J. Brizard, Rev. Mod. Phys. \textbf{81}, 693 (2009).
While we omit the energy of the drift velocity, we still have a drift velocity; 
the drift frequency (including all grad-B, curvature, and E$\times$B drifts) is given by
bounce-averaging the toroidal angular velocity, i.e.,
$\omega_d=\rmd{\theta}/\rmd t = {\partial H_{gc}}/{\partial\psi} $
In a homogeneous magnetic field, both $\omega_c$ and $\omega_b$ are constant, and then
$\omega_d$ evaluates the  E$\times$B drift frequency.


\bibitem{hbar}
To be correct about the dimension of $\mu$, 
$\mu/\hbar$ is the particle number, and $\hbar \omega_c$ is the energy level,
where $\hbar$ is a certain constant having the dimension of action
(which may not be Planck's constant in the classical regime).

\bibitem{YoshidaPPCF2013}
Z. Yoshida, H. Saitoh, Y. Yano, H. Mikami, N. Kasaoka, W. Sakamoto, J. Morikawa, M. Furukawa, and
S. M. Mahajan,
{Plasma Phys. Control. Fusion}
\textbf{55}, 014018 (2013).


\bibitem{hasegawa2005}
A. Hasegawa, Phys. Scr. {\bf T116}, 72 (2005).

\bibitem{Schippers}
P. Schippers, M. Moncuquet, N. Meyer-Vernet, and A. Lecacheux,
J. Geophys. Res. Space Phys. \textbf{118}, 7170 (2013). 

\bibitem{Grad1967}
H. Grad, Phys. Fluids \textbf{10}, 137 (1967).

\bibitem{Furukawa2014}
M. Furukawa, Phys. Plasmas \textbf{21}, 012511 (2014).

\bibitem{Gubar}
Yu. I. Gubar and A. R. Moszuhina,
Kosmicheskie issledovanija (Cosmic Research) \textbf{27}, 135 (1989).

\bibitem{yoshida2012}
Z. Yoshida,  Y. Yano, J. Morikawa, and H. Saitoh,
{Phys. Plasmas} \textbf{19}, 072303 (2012).

\bibitem{Weinstein}
A. Weinstein, Phys. Fluids \textbf{26}, 388 (1983).

\bibitem{Hasegawa-Mima}
A. Hasegawa and K. Mima, Phys. Rev. Lett. \textbf{39}, 205 (1977).

\bibitem{Charney}
J. C. Charney, J. Atmos. Sci. \textbf{28}, 1087 (1971).

\bibitem{Hasegawa1985} 
A. Hasegawa, Advances Phys. \textbf{34}, 1 (1985).

\bibitem{JBT}
J. B. Taylor,
Phys. Rev. Lett. \textbf{33}, 1139 (1974).

\bibitem{YoshidaMahajan2002}
Z. Yoshida and S. M. Mahajan,
Phys. Rev. Lett. {\bf 88},  095001 (2002)..

\bibitem{ItoYoshia1996}
N. Ito and Z. Yoshida,
Phys. Rev. E {\bf 53}, 5200  (1996).


\bibitem{dissipation}
One often assumes $\sigma$ to be equivalent to energy dissipation rate. 
The history of the latter concept goes back to Helmholtz' minimum dissipation principle, 
by which the velocity distribution of a slow stationary incompressible fluid can be determined; 
H. Helmholtz, Wiss. Abh \textbf{1}, 223 (1868).
Generalizing the notion of force and current (or flux), we may describe thermodynamic
non-equilibrium in parallel with mechanical non-equilibrium,
and then the mechanical minimum-dissipation principle (Helmholtz' principle) extends to the thermodynamic
minimum-dissipation principle (Onsager's principle); 
L. Onsager, Phys. Rev. \textbf{37}, 405 (1931); \textbf{38}, 2265 (1931).
Prigogine's minimum $\sigma$ principle has an independent origin; first it was applied to linear irreversible
processes in discontinuous systems; it was then extended to some nonlinear and
continuous systems; 
G. Nicolis and I. Prigogine, \textit{Self-organization in nonequilibrium systems --from dissipative
structures to order through fluctuations} (Wiley, New York, 1977).
For some physical systems, $\sigma$ becomes equivalent to the energy dissipation function; see 
I. Gyarmati, \textit{Nonequilibrium thermodynamics} (North-Holland, Amsterdam, 1962).





\bibitem{Sawada}
Y. Sawada, Prog. Theor. Phys. \textbf{66}, 68 (1981).

\bibitem{Ozawa}
H. Ozawa, S. Shimokawa, and H. Sakuma, 
Phys. Rev. E \textbf{64}, 026303 (2001).

\bibitem{YM2008}
Z. Yoshida and S. M. Mahajan,
Phys. Plasmas \textbf{15}, 032307 (2008).

\bibitem{Niven}
R. K. Niven,
Phys. Rev. E \textbf{80}, 021113 (2009).

\bibitem{RoderickDewar}
R. Dewar, J. Phys. A \textbf{36}, 631 (2003).

\bibitem{KawazuraYoshida2010}
Y. Kawazura and Z. Yoshida,
Phys. Rev. E {\bf 82}, 066403  (2010).

\bibitem{YoshidaKawazura2014}
Z. Yoshida and Y. Kawazura,
in \textit{Beyond The Second Law: Entropy Production and Non-Equilibrium Systems},
(Springer, New York, 2014), Chap. 15.



\end{thebibliography}
\end{document}